\documentclass[reprint,prl,aps,superscriptaddress,amsmath,amssymb]{revtex4-2}

\usepackage{graphicx} 
\usepackage{dcolumn}  
\usepackage{bm}       
\usepackage{hyperref} 
\hypersetup{
    colorlinks=true,
    linkcolor=blue,
    filecolor=magenta,      
    urlcolor=blue,
    citecolor=blue,
}
\usepackage{subcaption}
\usepackage{booktabs}
\usepackage{xcolor}
\usepackage{placeins}

\begin{document}

\title{Autonomous Emergence of Hamiltonian in Deep Generative Models}

\author{Wenjie Xi}
\email{wjxi@connect.hku.hk}
\affiliation{Department of Physics and HK Institute of Quantum Science \& Technology, The University of Hong Kong, Pokfulam Road, Hong Kong, China}
\affiliation{Department of Physics and Shenzhen Key Laboratory of Advanced Quantum Functional Materials and Devices, Southern University of Science and Technology, Shenzhen 518055, China}

\author{Wei-Qiang Chen}

\affiliation{Department of Physics and Shenzhen Key Laboratory of Advanced Quantum Functional Materials and Devices, Southern University of Science and Technology, Shenzhen 518055, China}

\date{\today}

\begin{abstract}
The unprecedented predictive success of deep generative models in complex many-body systems, such as AlphaFold 3, raises an epistemological question: do these networks merely memorize data distributions via high-dimensional interpolation, or do they autonomously deduce the underlying physical laws? 
To address this, we introduce a framework to extract the implicit physical interactions learned by generative models. 
Using the exact equivalence between the zero-noise limit of a diffusion score field and the thermodynamic restoring force, we directly compare the internal interaction structure of a trained neural network with the physical Hamiltonian. 
Applying this framework to a sequence-dependent, frustrated 1D $O(3)$ spin glass, we probe the latent representations of an $O(3)$-equivariant attention architecture trained solely on thermal equilibrium snapshots. 
Without imposing a locality cutoff or configuration independence, the dense scalar matrix in our architecture develops locality and nearly configuration-independent coefficients that recover the microscopic interactions of the spin glass with Pearson correlation $0.993$ on $2{,}000$ unseen sequences. This provides quantitative, falsifiable evidence for the \emph{emergence of the Hamiltonian} in a deep generative model.
Similarly, in an explicit-water phenol experiment, the network is trained only on coordinate data. With sufficient expressive capacity, a molecular-graph-guided Pairformer recovers the dominant bond, angle, and proper-torsion potential-of-mean-force (PMF) sectors. A sector-preserving projection of configuration-dependent coefficients onto constants yields $R_{\rm const}^2=0.9669$ on unseen molecular dynamics (MD), providing molecular-scale evidence for partial \emph{emergence of the Hamiltonian}.
\end{abstract}

\maketitle


Deep generative models, such as AlphaFold 3 \cite{af3} and diffusion models \cite{ho2020denoising, song2020score,park2025neural}, have achieved unprecedented success in predicting complex many-body conformations. 
This success raises a fundamental epistemological question: do these highly parameterized networks merely memorize data distributions via high-dimensional interpolation, or do they autonomously deduce the underlying microscopic physical laws \cite{Bahri2020,Iten2020Discovering,Udrescu2020AI}? 
To quantitatively verify this, one must reconstruct the governing Hamiltonians from equilibrium data, which is the central objective of inverse statistical mechanics (ISM) \cite{Jaynes1957, Nguyen2017inverse}. 
However, traditional ISM methods rely on maximum likelihood estimation and are severely bottlenecked by the intractable partition function \cite{ackley1985learning}. 
This computational barrier becomes formidable for systems with continuous geometric manifolds and frustrated energy landscapes. 
Consequently, it remains a profound challenge to provide falsifiable evidence regarding whether deep generative architectures genuinely internalize physical rules.

To bypass the intractable partition function, we leverage the framework of score matching \cite{hyvarinen2005estimation, vincent2011connection} and its profound connections to parameter estimation in physical models \cite{SohlDickstein2011}. 
In the zero-noise limit, the score field of a Riemannian diffusion model \cite{debortoli2022riemannian} is analytically equivalent to the thermodynamic restoring force, $\mathbf{s}_\theta \equiv -\beta \nabla^{\rm tan} H$ \cite{Arts2023,zaidi2022pre,holderrieth2024hamiltonian}. 
This exact equivalence provides a mathematical bridge for directly comparing the internal interaction structure of the frozen neural network with the physical Hamiltonian. 
Unlike the conventional PINN formulations \cite{Raissi2019PINN,Karniadakis2021PIML,Lu2021DeepXDE,Jagtap2020XPINN,Kharazmi2021hpVPINN}, whose loss functions contain strong- or weak-form residuals encoding \emph{a priori} physical knowledge, our models are trained with a DSM loss alone. In our framework, physical knowledge only restricts the admissible function space or basis.
Rather than evaluating full-atom proteins with unknown ground-truth interactions, we first apply this framework to a mathematically tractable yet physically complex benchmark: a sequence-dependent, frustrated 1D $O(3)$ spin glass \cite{Binder1986spin}.
The inherent geometrical frustration \cite{Ramirez1994mag} and quenched disorder in this system yield complex non-collinear ground states. This rugged energy landscape strictly precludes trivial statistical pattern matching.
However, a crucial question remains: does the generative model resort to complex high-dimensional interpolation, or does it genuinely discover the underlying microscopic physical laws?

To quantitatively address this question, we probe the latent representations of an $O(3)$-equivariant attention architecture \cite{villar2021scalars,satorras2021neural,Batzner2022E3} trained solely on thermal equilibrium snapshots of the spin glass. 
By applying classical invariant theory, we strictly preserve rotational equivariance and Newton's third law within the network. 
Without incorporating energetic labels, locality masks, or configuration-independence constraints, we directly read the frozen network's dense scalar matrix. Its out-of-distribution (OOD) coefficients recover the six microscopic interaction parameters with Pearson correlation $0.993$. Locality emerges as $84.97\%$ of the absolute weight in $W_\theta$ is concentrated at $r=1,2$, while the relative root-mean-square (RMS) coefficient variation across configurations is $4.04\%$. These observations show that the network recovers the interaction parameters and locality of the underlying Hamiltonian while developing approximate configuration independence.

We next apply this framework to the solvent-marginalized PMF of phenol in explicit water. With sufficient expressive capacity, a molecular-graph-guided Pairformer recovers the dominant bond, angle, and proper-torsion PMF sectors. A sector-preserving projection of its configuration-dependent coefficients onto constants yields $R_{\rm const}^2=0.9669$ on unseen MD, providing molecular-scale evidence for partial \emph{emergence of the Hamiltonian}.

Beyond addressing the epistemological question, this framework offers practical applications for evaluating deep generative models. 
First, it provides a systematic method to interpret highly parameterized architectures by explicitly extracting the learned microscopic interactions. 
Furthermore, direct coefficient readout provides a physics-informed criterion for validating OOD predictions. 
When a model generates novel conformations, its internal coefficients can be tested for locality, configuration stability, and consistency with a common interaction law. Large long-range coefficients or strong configuration dependence then provide an independent physical warning. 
Ultimately, this paradigm establishes a pathway to utilize deep generative architectures for discovering unknown effective many-body Hamiltonians directly from complex structural data.

\vspace{3mm}
\noindent \textbf{Physical Model and Theoretical Mapping.} --- 
To rigorously test whether deep generative models autonomously deduce underlying physical laws, we construct a mathematically tractable yet physically complex benchmark: a variable-length, one-dimensional (1D) chain of classical continuous spins $\mathbf{S}_i \in S^2$. 
To model sequence-dependent disorder, each site $i$ is assigned a binary identity $\tau_i \in \{A, B\}$. Crucially, this sequence-conditional formulation generates $2^N$ combinatorial variations of the microscopic Hamiltonian.
Coupled with variable-length sequences ($N \in [20, 50]$) and Open Boundary Conditions (OBC), this prevents the generative model from memorizing a single fixed-length absolute-position template or a single global energy landscape.

\begin{figure*}[htbp]
    \centering
    \includegraphics[width=0.95\textwidth]{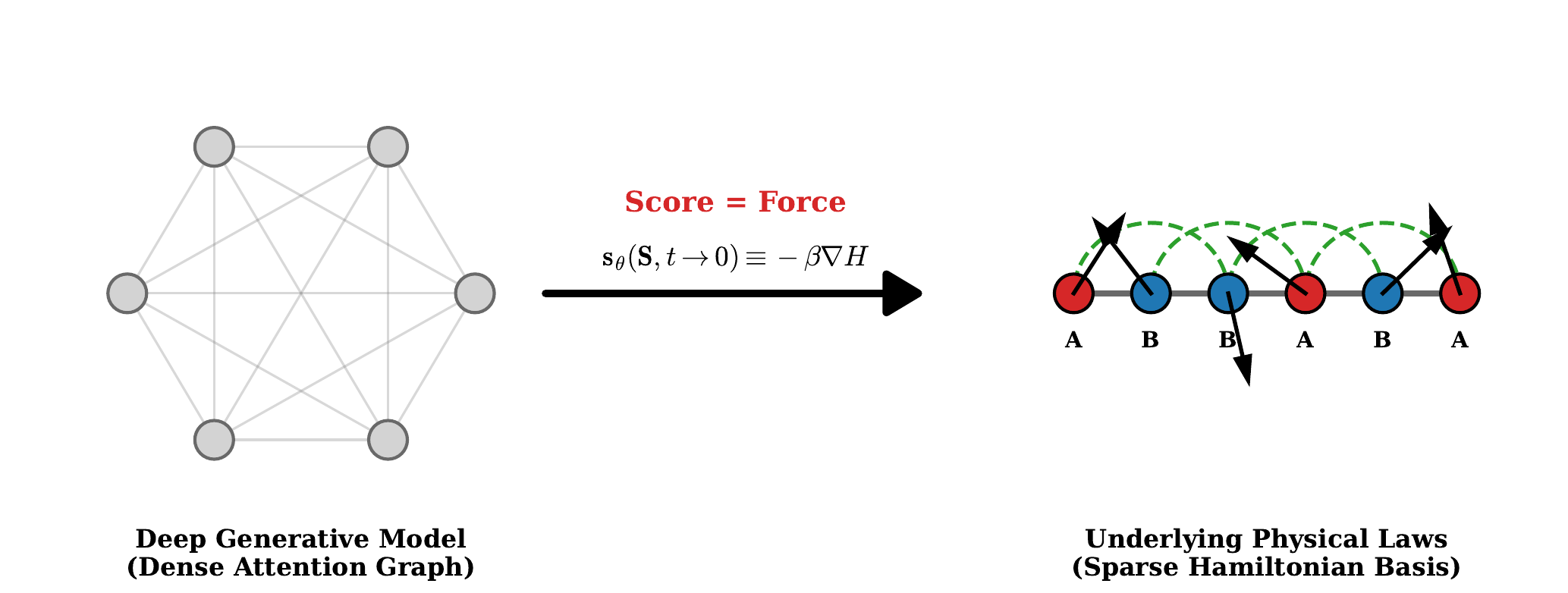}
    \caption{\textbf{Autonomous emergence of physical laws.} 
    \textbf{(Left)} Without locality, interaction-range, or Hamiltonian-coefficient priors, the deep generative model operates on a fully connected dense attention graph. 
    \textbf{(Center)} The exact equivalence between the zero-noise diffusion score field and the thermodynamic restoring force ($\mathbf{s}_\theta \equiv -\beta \nabla^{\rm tan} H$) allows the network to function as a direct force estimator. 
    \textbf{(Right)} Direct readout of the dense scalar matrix reveals its collapse onto sparse, sequence-dependent Hamiltonian coefficients. The model recovers the 1D spin-chain locality, sequence identities (A/B), and the frustrated nearest- and next-nearest-neighbor interactions.}
    \label{fig:concept}
\end{figure*}

The microscopic Hamiltonian governing the configurational energy landscape is defined by competing short-range (nearest-neighbor, $r=1$) and medium-range (next-nearest-neighbor, $r=2$) pairwise exchange interactions:
\begin{equation}
H(\mathbf{S}, \bm{\tau}) = -\sum_{i=1}^{N-1} J^{(1)}_{\tau_i \tau_{i+1}} (\mathbf{S}_i \cdot \mathbf{S}_{i+1}) - \sum_{i=1}^{N-2} J^{(2)}_{\tau_i \tau_{i+2}} (\mathbf{S}_i \cdot \mathbf{S}_{i+2})
\label{eq:hamiltonian}
\end{equation}

The underlying physics is entirely parameterized by exactly six fundamental coupling constants. 
We deliberately configure these parameters into a highly frustrated regime: all nearest-neighbor interactions are strictly ferromagnetic ($J^{(1)}_{AA}=1.0, J^{(1)}_{AB}=0.8, J^{(1)}_{BB}=0.5$), while all next-nearest-neighbor interactions are strictly antiferromagnetic ($J^{(2)}_{AA}=-0.45, J^{(2)}_{AB}=-0.35, J^{(2)}_{BB}=-0.25$).
This intense geometrical frustration, coupled with the quenched disorder of random sequences, forces the spins into non-collinear ground states, creating a rugged energy landscape that thwarts simple pattern matching.

To bypass the intractable partition function and probe the network, we leverage the exact isomorphism between continuous-state score matching and non-equilibrium thermodynamics. In the zero-noise limit, the generative score field analytically reduces to the microscopic thermodynamic restoring force: $\mathbf{s}_\theta(\mathbf{S}, t \to 0) \equiv -\beta \nabla_{\mathbf{S}}^{\rm tan} H(\mathbf{S}, \bm{\tau})$. To preserve the rigid $S^2$ geometric constraints of the spins, the diffusion process is formulated as spherical Brownian motion \cite{debortoli2022riemannian}, ensuring that the network learns a valid tangent vector field without injecting unphysical radial gradients (see Supplemental Material S2 for small-time Riemannian score targets and likelihood-weighted objectives). This equivalence allows us to utilize the converged score field as a direct mathematical conduit to extract the implicit Hamiltonian parameters (Fig.~\ref{fig:concept}).

\vspace{3mm}
\noindent \textbf{Equivariant Inference Architecture.} --- 
To act as a valid thermodynamic force estimator (see Supplemental Material S3 for deep learning architecture configurations), the neural network's predicted score field must strictly adhere to global $O(3)$ rotational equivariance. Mapping scalar neural features directly to a 3D vector fundamentally breaks this continuous symmetry. We therefore rely on the fundamental theorem of classical invariants \cite{villar2021scalars}, which states that an $O(3)$-equivariant vector function can be expressed as a linear combination of the input vectors, weighted by invariant scalar functions.

Consequently, our attention architecture processes rotationally invariant sequence and geometric features to predict a dense scalar matrix $\tilde{W}_{\theta,ij}(\mathbf{S},t,\bm{\tau})$. Because $W_{\theta,ij}$ receives the spin configuration $\mathbf{S}$, diffusion time $t$, and sequence $\bm{\tau}$ throughout the architecture, it may in general depend on all these elements. We impose Newton's third law through symmetrization and remove self-interactions by
\begin{equation}
W_{\theta,ij}=\frac{1}{2}\left(\tilde{W}_{\theta,ij}+\tilde{W}_{\theta,ji}\right),
\qquad W_{\theta,ii}=0,
\end{equation}
before aggregating the $O(3)$-equivariant vector $\mathbf{y}_i=\sum_j W_{\theta,ij}\mathbf{S}_j$. Finally, tangent projection gives the Riemannian score,
\begin{equation}
\mathbf{s}^{(i)}_\theta=\mathbf{P}_{\mathrm{tan}}(\mathbf{y}_i)
=\mathbf{y}_i-(\mathbf{y}_i\cdot\mathbf{S}_i)\mathbf{S}_i.
\end{equation}
All off-diagonal pairs remain active: no distance mask, interaction cutoff, or sparsity constraint is imposed on $W_\theta$. Nor is $W_\theta$ constrained to be configuration independent. We therefore test whether score training makes $W_\theta$ autonomously recover both the numerical coefficients and the sparse local structure of the Hamiltonian.

\begin{figure}[htbp]
    \centering
    \includegraphics[width=0.91\columnwidth]{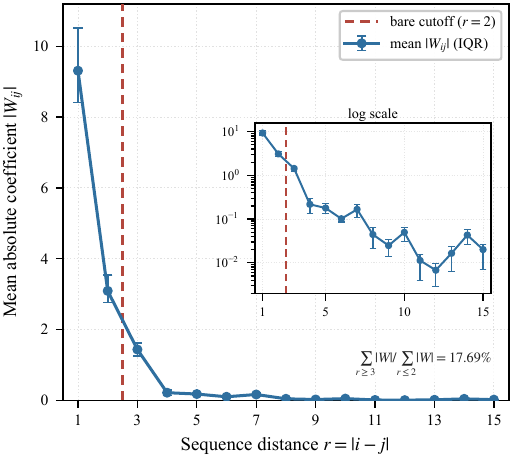}
    \caption{\textbf{Emergent locality of the dense scalar matrix.}
    The mean absolute coefficient $|W_{\theta,ij}|$ is evaluated over all $6{,}340{,}965$ off-diagonal pair observations from $10{,}000$ configurations of $2{,}000$ unseen sequences. It concentrates at $r=1,2$ and decays at longer range although the network is fully connected and has no locality cutoff. The absolute weight in $W_\theta$ over all $r\geq3$ is $17.69\%$ of that over $r=1,2$. Error bars denote the interquartile range; the inset shows the same profile on a logarithmic scale.}
    \label{fig:weight_decay}
\end{figure}

\vspace{3mm}
\noindent \textbf{Direct Scalar-Matrix Readout.} --- 
The generative model is trained solely on thermal equilibrium snapshots ($T_0=0.05$) generated via replica exchange Monte Carlo (see Supplemental Material S1).

We freeze this model and directly read $W_\theta$ at $t=10^{-4}$ on $2{,}000$ unseen sequences, with five saved equilibrium configurations. Equation~\ref{eq:hamiltonian} is used only to generate and evaluate the benchmark: standard denoising score matching (DSM) training receives no energetic labels or prior physical knowledge such as the $r=1,2$ support or the six coupling values. Here $\mathcal{J}(\bm{\tau})$ denotes the sequence-conditioned coupling parameters specified by Eq.~\ref{eq:hamiltonian}. On the $670{,}110$ interacting-pair observations, $W_\theta$ and $\mathcal{J}$ have Pearson correlation $0.9929$. The six class means define the class-balanced scale $a_\star=\frac16\sum_{c=1}^{6}\langle W\rangle_c/J_c=10.62$; the RMS dispersion of the six ratios $\langle W\rangle_c/J_c$ about $a_\star$ is $18.13\%$. Although every off-diagonal pair is available to the network, the absolute weight in $W_\theta$ over all noninteracting pairs ($r\geq3$) is $17.69\%$ of that over $r=1,2$. This simultaneous recovery of the interaction parameters and sparse local support constitutes the \emph{emergence of the Hamiltonian} at the scalar-matrix level.

\begin{figure}[t]
    \centering
    \includegraphics[width=\linewidth]{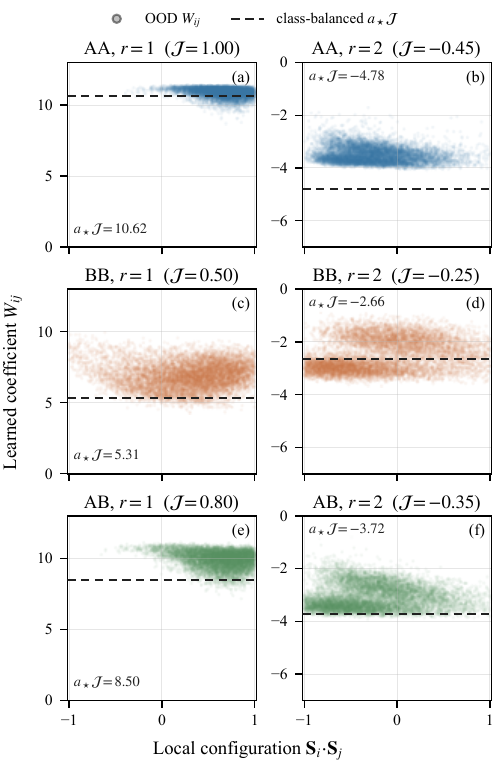}
    \caption{\textbf{Direct coefficient readout on unseen sequences.}
    Each panel shows the coefficient $W_{\theta,ij}$ for one of the six interaction parameters over OOD equilibrium configurations, plotted against the local configuration $\mathbf{S}_i\!\cdot\!\mathbf{S}_j$. The dashed line marks the constant $a_\star J$ obtained from the class-balanced global scale $a_\star=10.62$. The point clouds remain concentrated as the local configurations vary.}
    \label{fig:direct_wj}
\end{figure}
\vspace{3mm}
\noindent \textbf{Emergent Configuration Independence and Identifiability.} --- 
We see that the coefficients of the directly read dense scalar matrix change little as the configuration varies. For each unseen sequence and interacting pair, let $W_{\theta,ij}^{(k)}$ denote the coefficient read on its $k$th saved equilibrium configuration. The relative configuration RMS is
\begin{equation}
\sigma_{\rm conf}
=\left[
\frac{\mathbb{E}_{\bm{\tau},i,j}\operatorname{Var}_{k}
\!\left[W_{\theta,ij}^{(k)}\right]}
{\mathbb{E}_{\bm{\tau},i,j,k}\!\left[\left(W_{\theta,ij}^{(k)}\right)^2\right]}
\right]^{1/2}
=0.0404,\qquad r=1,2.
\label{eq:coefficient_collapse}
\end{equation}
Thus the configuration-dependent change is only $4.04\%$ of the RMS coefficient magnitude. No configuration average is used by the model or substituted into the score: every point in Fig.~\ref{fig:direct_wj} is the directly read $W_{\theta,ij}(\mathbf{S},t,\bm{\tau})$. The near-horizontal point clouds therefore show that approximate configuration independence emerges from training rather than from an architectural assumption.

We now give a concise identifiability argument for why low-noise score training can select one Hamiltonian law across the sequence ensemble. Write $H_{\bm{\eta}}=\sum_{a=1}^{6}\eta_a\Phi_a$, where the six $\Phi_a(\mathbf{S},\bm{\tau})$ represent the pair-energy sums for the different pair types and distances in Eq.~\ref{eq:hamiltonian}. Define the dimensionless couplings $\bm{\lambda}=\beta\bm{\eta}$ and tangent fields $\mathbf{B}_a=-\nabla_{\mathbf{S}}^{\rm tan}\Phi_a$. In the low-noise limit, a candidate Hamiltonian gives $\mathbf{s}_{\bm{\lambda}}=\sum_a\lambda_a\mathbf{B}_a$, while the data score is $\mathbf{s}_\star=\sum_a\lambda_{\star,a}\mathbf{B}_a$. The expected squared score loss therefore becomes
\begin{equation}
\begin{aligned}
\mathcal{L}(\bm{\lambda})
&=\mathbb{E}\left\|\mathbf{s}_{\bm{\lambda}}-\mathbf{s}_\star\right\|^2 \\
&=\mathbb{E}\left\|\sum_a(\lambda_a-\lambda_{\star,a})\mathbf{B}_a\right\|^2 \\
&=\sum_{a,b}(\lambda_a-\lambda_{\star,a})(\lambda_b-\lambda_{\star,b})
\,\mathbb{E}\langle\mathbf{B}_a,\mathbf{B}_b\rangle \\
&=(\bm{\lambda}-\bm{\lambda}_\star)^{\mathsf T}
G(\bm{\lambda}-\bm{\lambda}_\star),
\qquad G_{ab}=\mathbb{E}\langle\mathbf{B}_a,\mathbf{B}_b\rangle .
\end{aligned}
\label{eq:pooled_identifiability}
\end{equation}
For the sampled sequence and configuration ensemble, the empirical pooled Gram matrix of the six fields $\mathbf{B}_a$ is positive definite, indicating that Eq.~\ref{eq:pooled_identifiability} has the unique minimum $\bm{\lambda}_\star$ within this six-coupling Hamiltonian family. Thus pooled low-noise score matching selects a common interaction-parameter law across sequences, providing the identifiability mechanism for the observed \emph{emergence of the Hamiltonian}.

\vspace{3mm}
\noindent \textbf{Molecular-Graph-Guided PMF Emergence.} ---
To test whether this framework extends to a realistic molecular system, we consider 13-atom phenol solvated by 512 water molecules. Let $\mathbf R=(\mathbf r_1,\ldots,\mathbf r_{13})$ and $\mathbf X$ denote the solute and solvent coordinates. Here $\beta=(k_{\rm B}T)^{-1}$, and $H(\mathbf R,\mathbf X)$ is the Hamiltonian. Integrating over the solvent coordinates gives the dimensionless PMF action
\begin{equation}
\begin{aligned}
S_{\rm PMF}(\mathbf R)&=-\log\!\int d\mathbf X\,
e^{-\beta H(\mathbf R,\mathbf X)}+C,\\
\mathbf s_\star(\mathbf R)&=-\nabla_{\mathbf R}S_{\rm PMF}
=\beta\,\mathbb E[\mathbf F_{\mathbf R}\mid\mathbf R],\\
\mathbf F_{\mathbf R}&\equiv-\nabla_{\mathbf R}H(\mathbf R,\mathbf X).
\end{aligned}
\label{eq:phenol_pmf}
\end{equation}
Here $C$ is an arbitrary additive constant.
All three networks are trained with the same coordinate-only DSM objective. We write $\mathbf s^{(m)}_\sigma=-\nabla_{\mathbf R}S^{(m)}_\sigma$ for the score of Network $m$ at Gaussian noise scale $\sigma$. Their function spaces are
\begin{equation}
\begin{aligned}
u_{c\mu k}(\mathbf R)&=\sum_{e\in\mathcal E_{c\mu}}
b_{c\mu k}\!\left(q_e(\mathbf R)\right),\\
\Phi_a(\mathbf R)&=\sum_{\mu k}T^{(c)}_{a,\mu k}
u_{c\mu k}(\mathbf R),\quad a\in\mathcal I_c,\\
S^{(1)}_\sigma(\mathbf R)&=\mathcal F_{\theta_1}
\!\left(\{Z_i\},\{r_{ij}\},\sigma\right),\\
S^{(2)}_\sigma(\mathbf R)&=\bm\alpha_\sigma^{\mathsf T}\bm\Phi(\mathbf R),\\
S^{(3)}_\sigma(\mathbf R)&=\mathcal F_{\theta_3}\!\left(\bm\Phi(\mathbf R),\sigma\right),\\
\mathbf s^{(3)}_\sigma(\mathbf R)&=-\sum_{a=1}^{240}
w^{(3)}_{\sigma a}(\mathbf R)\nabla_{\mathbf R}\Phi_a(\mathbf R),
\quad w^{(3)}_{\sigma a}=\frac{\partial\mathcal F_{\theta_3}}{\partial\Phi_a}.
\end{aligned}
\label{eq:phenol_spaces}
\end{equation}
Here $c$ labels one of six graph-defined geometric sectors (bond, angle, proper torsion, out-of-plane, graph-distance-3 pair, or longer-range pair). $e$ indexes one actual event, such as a specified bond between two graph-adjacent atoms or an angle formed by three consecutive atoms. $\mu$ labels an event type whose events share the same atomic species and local graph environment, such as aromatic C--C bonds or C--C--C ring angles. The scalar $q_e$ is the length, angle, dihedral, squared scalar triple product of three normalized bond vectors, or pair distance of event $e$, and $b_{c\mu k}$ is the $k$th generic radial, angular, or Fourier basis function for event type $\mu$ in sector $c$. The fixed linear transformation $T^{(c)}$ combines $u_{c\mu k}$ into linearly independent basis functions $\Phi_a$. The set $\mathcal I_c$ contains the indices of the basis functions belonging to sector $c$, and $\bm\Phi=(\Phi_a)_{a=1}^{240}$. Network 1 is a graph-free scalar Pairformer of atomic-species labels $Z_i$ and distances $r_{ij}=|\mathbf r_i-\mathbf r_j|$. Network 2 imposes configuration-independent coefficients $\bm\alpha_\sigma$, whereas Network 3 allows $\bm w^{(3)}_\sigma(\mathbf R)$ to vary with configuration.

All three networks are trained on 5000 configurations from 10 ns of MD in five independent water environments. Model selection uses 1000 configurations from a sixth independent 2-ns trajectory. The frozen networks are tested on 1200 configurations from 24 ns of MD in eight new water environments.

For each structured network $m=2,3$ and each sector $c$, we directly read the internal score assigned by the trained network, $\widehat{\mathbf s}^{(m)}_c=-\sum_{a\in\mathcal I_c}\widehat w^{(m)}_a\nabla_{\mathbf R}\Phi_a$, with $\widehat{\bm w}^{(2)}=\widehat{\bm\alpha}$ and $\widehat{\bm w}^{(3)}=\partial\mathcal F_{\theta_3}/\partial\bm\Phi$. A hat denotes evaluation of the frozen network at the common lowest noise, $\sigma=0.000375$ nm. On the independent test configurations, we compare this score with $\mathbf f_c=\beta\mathcal P_{\rm int}\mathbf F^{\rm OMM}_c$, where $\mathbf F^{\rm OMM}_c$ is the aligned OpenMM force component and $\mathcal P_{\rm int}$ removes global translation and rotation. The six model sectors are deliberately aligned with the corresponding OpenMM force-field components, but no component energies, forces, or parameters enter training. For each network $m$, Table~\ref{tab:phenol_pmf} reports the force cosine and normalized root-mean-square error (NRMSE), defined by
\begin{equation}
\begin{aligned}
 {\rm cos}^{(m)}_c&=\frac{\langle\widehat{\mathbf s}^{(m)}_c\!\cdot\!\mathbf f_c\rangle}{\sqrt{\langle\|\widehat{\mathbf s}^{(m)}_c\|^2\rangle\langle\|\mathbf f_c\|^2\rangle}},\\
 {\rm NRMSE}^{(m)}_c&=\sqrt{\frac{\langle\|\widehat{\mathbf s}^{(m)}_c-\mathbf f_c\|^2\rangle}{\langle\|\mathbf f_c\|^2\rangle}}.
\end{aligned}
\label{eq:phenol_metrics}
\end{equation}
where the brackets average over all 1200 test configurations and 13 solute atoms.

Network 1 reaches total-force cosine $0.961$, showing that the graph-free Pairformer has sufficient expressive capacity for the total PMF score. Network 2 raises this to $0.992$ and learns its dominant bond, angle, and proper-torsion components accurately (Table~\ref{tab:phenol_pmf}), although the out-of-plane and pair sectors remain weak. Network 3 retains a total cosine $0.967$ and substantial bond, angle, and proper-torsion PMF readouts while allowing its coefficients to depend on configuration.

To test whether each learned sector approaches a configuration-independent Hamiltonian term, we project each frozen low-noise training score $\widehat{\mathbf s}^{(3)}_c$ separately:
\nopagebreak[4]
\begin{equation}
\begin{aligned}
\mathbf s_{\bm\alpha_c,c}&\equiv-
\sum_{a\in\mathcal I_c}\alpha_{ca}\nabla_{\mathbf R}\Phi_a,\\
\bm\alpha_c^\star&=\arg\min_{\bm\alpha_c}
\left\langle\left\|\widehat{\mathbf s}^{(3)}_c-
\mathbf s_{\bm\alpha_c,c}\right\|^2\right\rangle_{\rm train}
+\lambda_c\|\bm\alpha_c\|^2,\\
\mathbf s^{(3\text{-P})}_c&=\mathbf s_{\bm\alpha_c^\star,c},
\qquad
\mathbf s^{(3\text{-P})}=\sum_c\mathbf s^{(3\text{-P})}_c,\\
R^2_{\rm const}&=1-
\frac{\langle\|\widehat{\mathbf s}^{(3)}-\mathbf s^{(3\text{-P})}\|^2\rangle_{\rm OOD}}
{\langle\|\widehat{\mathbf s}^{(3)}\|^2\rangle_{\rm OOD}}\\
&=0.9669.
\end{aligned}
\label{eq:phenol_collapse}
\end{equation}
For each candidate $\lambda_c$, the constant coefficients $\boldsymbol\alpha_c$ are fitted on the five training families. We select $\lambda_c$ on the sixth family by minimizing the reconstruction error between the projected sector score and the corresponding frozen Network 3 score.

\begin{table}[t]
\caption{\textbf{Phenol PMF force-score recovery on independent MD at the common lowest noise.} Network 3-P is the sector-preserving constant projection of Network 3 defined in Eq.~\eqref{eq:phenol_collapse}. Entries are cosine/NRMSE; N/A means that Network 1 has no sector allocation.}
\label{tab:phenol_pmf}
\centering
\tiny
\begin{ruledtabular}
\setlength{\tabcolsep}{0.8pt}
\begin{tabular}{lcccc}
Readout & Net. 1 & Net. 2 & Net. 3 & Net. 3-P \\
\hline
Total PMF & $0.961/0.283$ & $0.992/0.137$ & $0.967/0.258$ & $0.982/0.209$ \\
Bond & N/A & $0.965/0.275$ & $0.943/0.346$ & $0.956/0.321$ \\
Angle & N/A & $0.978/0.211$ & $0.954/0.312$ & $0.976/0.255$ \\
Proper torsion & N/A & $0.993/0.275$ & $0.929/0.609$ & $0.971/0.592$ \\
Out-of-plane & N/A & $0.684/1.166$ & $0.766/2.874$ & $0.843/2.350$ \\
Graph-distance-3 pair & N/A & $0.177/1.020$ & $0.066/1.154$ & $0.085/1.100$ \\
Longer-range pair & N/A & $0.410/1.194$ & $-0.020/2.288$ & $-0.024/2.070$ \\
\end{tabular}
\end{ruledtabular}
\end{table}
\FloatBarrier

The sector-preserving constant projection, denoted Network 3-P, achieves $R^2_{\rm const}=0.9669$, raises the total force-score cosine to $0.982$, and improves the bond, angle, proper-torsion, and out-of-plane readouts. Thus most of the learned configuration dependence is unnecessary. This is evidence for partial \emph{emergence of the Hamiltonian} and is qualitatively consistent with the emergent Hamiltonian of the spin glass. However, gauge redundancy leading to component non-identifiability remains, as seen most clearly in the weak pair sectors and the out-of-plane amplitude.

\noindent \textbf{Conclusion.} --- 
For the spin glass, our direct scalar-matrix readout and pooled identifiability analysis provide falsifiable evidence that a deep generative model trained solely on equilibrium configurations approximately recovers the sparse interaction structure and relative microscopic coefficients as a shared law.

For a realistic molecule in explicit water, Network 1 establishes Pairformer capacity. Network 2 resolves the dominant bond, angle, and proper-torsion PMF components in a molecular-graph-guided basis. Network 3 retains substantial component readouts without a configuration-independence constraint. Network 3-P shows that most of these learned component scores can be represented by configuration-independent coefficients. However, for realistic complex molecules, simultaneously retaining sufficient expressive capacity and eliminating redundant gauge freedoms during Hamiltonian learning remains an important challenge.

\textbf{Acknowledgments:} This work was supported by Research Grants Council of Hong Kong (CRF C7015-24G, GRF 17311322 and CRF C7012-21GF), National Natural Science Foundation of China (Grant No. 12222416),
National Key R$\&$D Program of China (Grants No. 2022YFA1403700), NSFC (Grants No. 12141402) and the Science, Technology and Innovation Commission of Shenzhen Municipality (No. ZDSYS20190902092905285).

\appendix

\section{S1. Dataset Generation and Replica-Exchange Monte Carlo (REMC) Audit}

To construct the benchmark dataset of the frustrated 1D $O(3)$ spin glass, we generated $40,000$ distinct training sequences and $2,000$ out-of-sample testing sequences, with zero overlap between the training and testing sequences. The sequence lengths $N$ were drawn from the discrete uniform distribution on $\{20,\ldots,50\}$, and each site's identity $\tau_i \in \{A, B\}$ was assigned uniformly at random. 

Due to the intense geometrical frustration and quenched disorder, the energy landscape is exceedingly rugged. Standard Metropolis-Hastings algorithms suffer from severe critical slowing down and easily become trapped in metastable non-collinear valleys. 
To explore widely separated metastable regions at the target temperature $T_0=0.05$, we employed an accelerated REMC algorithm \cite{Hukushima1996}.

Specifically, $M=16$ replicas were simulated simultaneously at temperatures geometrically spaced between $T_{\min} = 0.05$ and $T_{\max} = 5.0$. One sweep comprises one attempted update of every spin in every replica. Adjacent replicas $m$ and $m+1$ attempted to exchange their configurations every 10 sweeps with probability $\min\{1,\exp[(\beta_m-\beta_{m+1})(E_m-E_{m+1})]\}$.

Within each replica, local spin updates were performed using a highly efficient hybrid strategy. 
For each site $i$, the local effective field is defined as $\mathbf{H}_{\text{eff}, i} = -\nabla_{\mathbf{S}_i} H(\mathbf{S}, \bm{\tau})$. We mixed two distinct update mechanisms with equal probability (50\%/50\%):
\begin{enumerate}
    \item \textbf{von Mises-Fisher Heat Bath:} The spin $\mathbf{S}_i$ is resampled directly from the exact local conditional Boltzmann distribution, which on the $S^2$ manifold corresponds to the von Mises-Fisher distribution: $P(\mathbf{S}_i) \propto \exp(\beta \mathbf{H}_{\text{eff}, i} \cdot \mathbf{S}_i)$. This step ensures thermalization and energy exchange with the heat bath.
    \item \textbf{Microcanonical Over-Relaxation \cite{Brown1987,Creutz1987}:} The spin $\mathbf{S}_i$ is deterministically reflected across the axis of the local effective field: 
    \begin{equation}
    \mathbf{S}_i \leftarrow 2(\mathbf{S}_i \cdot \hat{\mathbf{u}})\hat{\mathbf{u}} - \mathbf{S}_i,
    \end{equation}
    where $\hat{\mathbf{u}} = \mathbf{H}_{\text{eff}, i} / \|\mathbf{H}_{\text{eff}, i}\|$. This geometric operation strictly conserves the local energy ($\Delta E = 0$) while producing a large move in spin orientation, thereby accelerating decorrelation without rejection.
\end{enumerate}
For each sequence, we discarded the first $2{,}000$ sweeps and then saved configurations at $T_0$ every $100$ sweeps.

\begin{figure*}[htbp]
    \centering
    \includegraphics[width=0.98\textwidth]{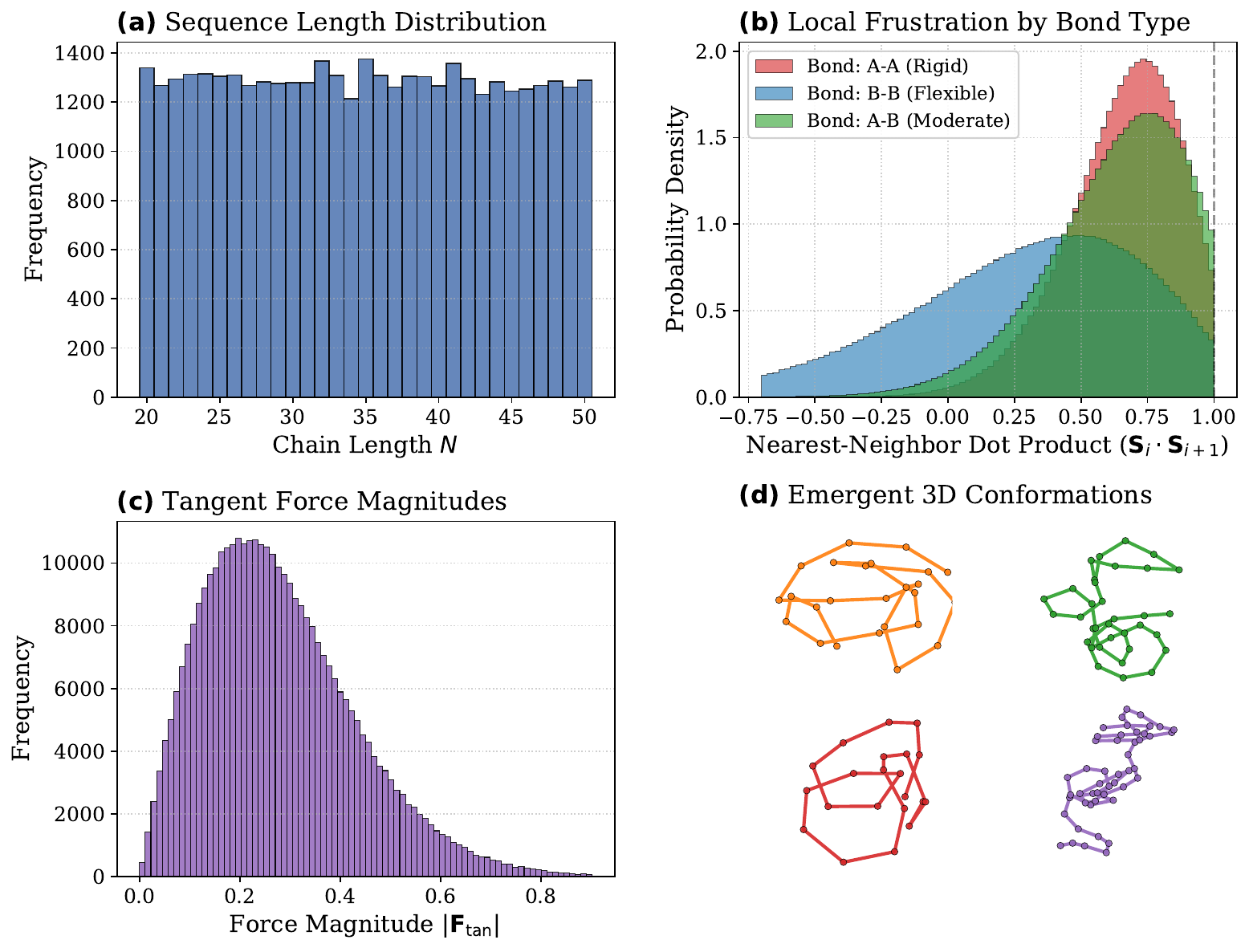}
    \caption{\textbf{Statistical and geometric properties of the frustrated 1D $\mathbf{O(3)}$ spin glass dataset.} 
    \textbf{(a)} The uniform sequence length distribution ($N \in[20, 50]$) prevents memorization of a single fixed-length positional template. 
    \textbf{(b)} Sequence-dependent local frustration. 
    The probability density of nearest-neighbor inner products bifurcates based on bond identity. 
    Strong $J^{(1)}_{AA}$ couplings enforce rigid ferromagnetic alignments (peak near $+0.75$), while weaker $J^{(1)}_{BB}$ couplings act as flexible hinges to accommodate geometrical frustration. 
    \textbf{(c)} Distribution of theoretical tangent force magnitudes ($|\mathbf{F}_{\text{tan}}|$), where $\mathbf F_{\rm tan}=-\nabla_{\mathbf S}^{\rm tan}H$. The smooth envelope is bounded away from zero, ensuring a well-defined, non-singular gradient field for diffusion model training. 
    \textbf{(d)} Integrating the 1D continuous spins into 3D spatial coordinates yields compact, non-collinear folded conformations, reflecting the structural heterogeneity induced by geometrical frustration and quenched disorder.}
    \label{fig:dataset_overview}
\end{figure*}

We tested configurational decorrelation using a separately initialized REMC run for $2{,}000$ of the same sequences. To remove the exact global $O(3)$ symmetry, we define the aligned overlap $q(\mathbf S,\mathbf S')=\max_{R\in O(3)}N^{-1}\sum_i\mathbf S_i\!\cdot R\mathbf S'_i$. The mean overlap between configurations separated by $100$ sweeps was $0.5095\,[0.5062,0.5126]$, statistically indistinguishable from the independent-run baseline $0.5081\,[0.5051,0.5111]$. Here brackets denote $95\%$ sequence-bootstrap intervals, obtained by resampling sequences with replacement. We therefore detect no metastable locking for the saved configurations.
The global statistical and geometric properties of the ensemble are summarized in Fig.~\ref{fig:dataset_overview}.

\section{S2. Riemannian Score Target and Likelihood Weighting}

Injecting Euclidean Gaussian noise into $S^2$ spins violates the rigid manifold constraint $|\mathbf{S}_i|=1$. Consequently, the forward diffusion process is modeled as spherical Brownian motion. 
According to the small-time asymptotic expansion of the heat kernel on Riemannian manifolds, the leading conditional-score target pointing from the perturbed state $\mathbf{S}_t$ back to the clean state $\mathbf{S}_0$ along the shortest geodesic is given by the Riemannian logarithm map:
\begin{equation}
\mathbf{U}_{\text{target}}(\mathbf{S}_t, \mathbf{S}_0) = \frac{1}{t} \log_{\mathbf{S}_t}(\mathbf{S}_0) = \frac{\gamma}{t \sin\gamma} \Big( \mathbf{S}_0 - \cos\gamma\, \mathbf{S}_t \Big),
\label{eq:riemannian_log}
\end{equation}
where $\gamma = \arccos(\mathbf{S}_0 \cdot \mathbf{S}_t) \in [0, \pi]$ is the geodesic distance. 
For $\gamma<10^{-4}$, we replace $\gamma/\sin\gamma$ by its limit, 1.

A critical challenge in training diffusion models with Eq.~\ref{eq:riemannian_log} is that the target variance scales as $\mathcal{O}(1/t)$ when $t \to 0$, causing gradient explosion during standard mean-squared-error (MSE) minimization. 
To stabilize training and preserve the thermodynamic restoring-force limit, we employ likelihood weighting. The loss function is explicitly multiplied by the diffusion time $t$:
\begin{equation}
\begin{aligned}
\mathcal{L}(\theta) = \mathbb{E}_{t, \mathbf{S}_0, \mathbf{S}_t}\!\Bigg[
\sum_{i=1}^{N} M_i t\, \big\|
&\mathbf{s}_\theta^{(i)}(\mathbf{S}_t,t,\bm{\tau}) \\
&-\mathbf{U}_{\text{target}}^{(i)}(\mathbf{S}_t,\mathbf{S}_0)
\big\|^2 \Bigg] .
\end{aligned}
\end{equation}
During the later stage of training, approximately $25\%$ of diffusion-time samples satisfy $t<10^{-3}$. The binary mask $M_i \in \{0, 1\}$ ensures that the loss is only computed on valid spins, neutralizing arbitrary contributions from padded sequences in batched computations. This $t$-weighted scaling stabilizes the gradients and strictly aligns the objective with the expected log-likelihood bound.

\section{S3. Neural Architecture Configuration}

Our continuous-state generative model adapts the Pairformer architecture to respect $O(3)$ spatial symmetries. 
The model was implemented in PyTorch and explicitly decouples sequence representations from pairwise geometric features.

\textbf{Representations:} The single-node representation $\mathbf{h}_i \in \mathbb{R}^{64}$ embeds the binary sequence identity (A or B), an absolute 1D positional encoding, and a continuous time embedding obtained by mapping the scalar $t$ through a multilayer perceptron (MLP). 
The pair representation $\mathbf{z}_{ij} \in \mathbb{R}^{32}$ embeds the sequence separation distance $|i-j|$. Crucially, to maintain strict $O(3)$ rotational invariance, spatial geometry is only fed into $\mathbf{z}_{ij}$ through the scalar inner product matrix $(\mathbf{S}_i \cdot \mathbf{S}_j)$, rather than Cartesian coordinate differences.

\textbf{Attention Blocks:} The network iteratively refines $\mathbf{h}_i$ and $\mathbf{z}_{ij}$ via four Pairformer blocks. 
Each block employs a four-head attention mechanism in which the single-node query, key, and value matrices are biased by the pair representations. 
The attention logits are rigorously masked (set to $-10^9$) to prevent any information leakage between valid spin sites and padding tokens used for batch alignment. 

\textbf{Invariant Output Head:} The final stage of the network transforms the abstract, high-dimensional neural features into a physically valid thermodynamic force. 
This process involves three explicit steps to enforce physical and engineering constraints:
\begin{enumerate}
    \item \textit{Scalar Projection:} First, the 32-dimensional pair representation $\mathbf{z}_{ij}$ (encoding the learned interactions between site $i$ and $j$) is projected to a single scalar value via an MLP block ($\mathbb{R}^{32} \to \mathbb{R}^{32} \to \text{GELU} \to \mathbb{R}^1$). This produces a raw, asymmetric weight matrix $\tilde{W}_{ij}$, which acts as the unconstrained interaction strength from spin $i$ to spin $j$.
    \item \textit{Newton's Third Law (Symmetrization):} A raw neural network output does not inherently respect physical conservation laws; the predicted force exerted by $i$ on $j$ might differ from that exerted by $j$ on $i$. 
    To strictly enforce Newton's third law of action and reaction, we explicitly symmetrize the matrix. Furthermore, we zero out the diagonal to preclude any unphysical self-interactions (a spin cannot exert a force on itself). 
    This yields the physically valid interaction matrix: $W_{ij} = (\tilde{W}_{ij} + \tilde{W}_{ji})/2$, with $W_{ii}=0$.
    \item \textit{2D Masking of Padding Artifacts:} Because variable-length sequences are padded to a maximum length for batched GPU computations, we must prevent ``ghost'' padded spins from exerting or receiving forces. We construct a 2D physical mask $M_{ij}=M_iM_j\in\{0,1\}^{N\times N}$ from the 1D valid-spin mask. Multiplying $W_{ij}$ by this 2D mask sets any interaction involving a padded token to zero.
\end{enumerate}
\section{S4. Global Energy Scale for Direct Coefficient Readout}

Equilibrium coordinate data determine the dimensionless couplings $\beta J$, but not the bare energy scale $J$ and inverse temperature $\beta$ separately. The raw pair-head coefficients may additionally contain one global normalization associated with the diffusion parameterization. We therefore compare $W_\theta$ and $\mathcal{J}$ up to a single scalar $a_\star$.

Let $\mu_c=\langle W_{\theta,ij}\rangle_c$ be the empirical mean for coupling class $c$ over the $2{,}000$ unseen sequences and five configurations per sequence. Because the physical object contains six coupling constants, we give the six nonzero classes equal weight and define
\begin{equation}
a_\star=\underset{a}{\arg\min}\;\frac{1}{6}\sum_{c=1}^{6}
\left(\frac{\mu_c}{J_c}-a\right)^2
=\frac{1}{6}\sum_{c=1}^{6}\frac{\mu_c}{J_c}
=10.6244.
\end{equation}
The relative RMS dispersion of the six ratios is
\begin{equation}
\epsilon_{\rm class}
=\left[\frac{1}{6}\sum_{c=1}^{6}
\left(\frac{\mu_c/J_c}{a_\star}-1\right)^2\right]^{1/2}
=0.1813.
\end{equation}
Under this fixed scale, the pointwise relative RMSE is $\epsilon_{\rm point}=\{\langle[W_{\theta,ij}-a_\star J_{c(i,j)}]^2\rangle_{\rm supp}/\langle W_{\theta,ij}^2\rangle_{\rm supp}\}^{1/2}=0.1613$, where $c(i,j)$ is the coupling class of an interacting pair and the averages run over all $r=1,2$ observations. The Pearson correlation is $0.9929$.

\section{S5. Explicit-Water Phenol PMF Experiment}

The system contains 13 phenol atoms and 512 explicit amber14-TIP3P water molecules in a 2.50-nm periodic box, with no dummy atoms. Production used constant number of particles, volume, and temperature (NVT) dynamics at 298.15 K, a 2-fs time step, and particle-mesh Ewald electrostatics. We call each independently initialized MD trajectory a family; each family used its own solvent configuration and velocity seed. Only the periodically unwrapped and centered phenol coordinates entered DSM training. Water coordinates, energies, forces, and force-field parameters were excluded from training. The phenol atoms are unconstrained, so their centered Cartesian coordinates form a Euclidean translation-free space. We therefore use centered Cartesian Gaussian perturbations.

The fit set contains 5000 configurations from five independent families, corresponding to 10 ns of production, and model selection uses 1000 configurations from a sixth 2-ns family. The holdout comprises eight newly initialized families, each with 1 ns equilibration and 3 ns production. We selected 150 configurations per family at 20-ps spacing, giving 1200 configurations from 24 ns of total production.

All three networks use the same fit/selection protocol, centered antithetic DSM perturbations, and noise scales $\sigma=(0.000375,0.00075,0.0015)$ nm. Network 1 is a graph-free scalar Pairformer of atomic-species labels and all pair distances. Network 2 is linear in 240 graph-conditioned basis functions with configuration-independent coefficients. Network 3 applies a scalar Pairformer to the same basis functions, so its effective coefficients may depend on configuration. The initial basis set contains 270 generic event sums across the bond-radial, angle, proper-torsion, out-of-plane, graph-distance-3 pair, and longer-range pair sectors; within-sector score-Gram truncation leaves 240 independent combinations.

For each sector--type pair $(c,\mu)$, let $Q_p^{c\mu}$ be the $p$th percentile of $q_e(\mathbf R_n)$ pooled over the 5000-configuration fit set $\mathbf R_n\in\mathcal D_{\rm fit}$ and events $e\in\mathcal E_{c\mu}$. For the five sectors $c\ne\mathrm{proper}$, the single-event basis functions $b_{c\mu k}$ used in $u_{c\mu k}$ in Eq.~\eqref{eq:phenol_spaces} are
\begin{equation}
\begin{aligned}
m_{c\mu}&=Q_{50}^{c\mu},\qquad
\ell_{c\mu}=\frac{Q_{75}^{c\mu}-Q_{25}^{c\mu}}{1.349},\\[-2pt]
z_e&=\frac{q_e-m_{c\mu}}{\ell_{c\mu}},\\[-2pt]
g_\xi(z)&=\exp\!\left[-\frac{(z-\xi)^2}{2h^2}\right],\\[-2pt]
\{b_{c\mu k}(q_e)\}_k
&=\{z_e,z_e^2,g_{-2}(z_e),\ldots,g_2(z_e)\},\\[-2pt]
\{b_{{\rm proper}\,\mu k}(\phi_e)\}_k
&=\{\sin n\phi_e,\cos n\phi_e\}_{n=1}^{3}.
\end{aligned}
\end{equation}
Here $m_{c\mu}$ is the median, $\ell_{c\mu}$ the Gaussian-calibrated interquartile scale, and $h=0.85$. For $c\ne\mathrm{proper}$, $q_e$ is a bond length, bond angle, squared scalar triple product of three normalized bond vectors, or pair distance, as appropriate. For an out-of-plane event $e=(i;j,k,l)$ centered at atom $i$, this coordinate is $q_e=[\widehat{\mathbf r}_{ij}\cdot(\widehat{\mathbf r}_{ik}\times\widehat{\mathbf r}_{il})]^2$, where $\widehat{\mathbf r}_{ij}=(\mathbf r_j-\mathbf r_i)/|\mathbf r_j-\mathbf r_i|$. For a proper-torsion event $e=(i,j,k,l)$, $\phi_e\in(-\pi,\pi]$ is the signed dihedral angle between the $(i,j,k)$ and $(j,k,l)$ planes.

\renewcommand{\bibfont}{\fontsize{6.5}{6.8}\selectfont}
\bibliographystyle{apsrev4-2} 
\bibliography{ref}

\end{document}